\documentclass{aip-cp}

\usepackage[numbers]{natbib}
\usepackage{rotating}
\usepackage{graphicx}
\usepackage{url}

\begin{document}

\title{Abundance Uncertainties Obtained With the \textsc{PizBuin} Framework For Monte Carlo Reaction Rate Variations}

\author[aff1,aff2]{T. Rauscher\corref{cor1}\noteref{note1}}
\author[aff3,aff4]{N. Nishimura\corref{cor1}}
\author[aff1,aff5]{G. Cescutti\corref{cor1}}
\author[aff3,aff6]{R. Hirschi\corref{cor1}}
\author[aff7]{A. St.J. Murphy\corref{cor1}}

\affil[aff1]{Centre for Astrophysics Research, University of Hertfordshire, Hatfield AL10 9AB, United Kingdom}
\affil[aff2]{Department of Physics, University of Basel, 4056 Basel, Switzerland}
\affil[aff3]{Astrophysics group, Lennard-Jones Laboratories, Keele University, Keele ST5 5BG, United Kingdom}
\affil[aff4]{Yukawa Institute for Theoretical Physics, Kyoto University, Kyoto 606-8502, Japan}
\affil[aff5]{INAF, Osservatorio Astronomico di Trieste, 34131 Trieste, Italy}
\affil[aff6]{Kavli IPMU (WPI), University of Tokyo, Kashiwa 277-8583, Japan}
\affil[aff7]{School of Physics and Astronomy, University of Edinburgh, Edinburgh EH9 3FD, United Kingdom}
\corresp[cor1]{UK Network for Bridging Disciplines of Galactic Chemical Evolution (BRIDGCE), \url{https://www.bridgce.net}}
\authornote[note1]{Corresponding author}

\maketitle

\begin{abstract}
Uncertainties in nucleosynthesis models originating from uncertainties in astrophysical reaction rates were estimated in a Monte Carlo variation procedure. Thousands of rates were simultaneously varied within individual, temperature-dependent errors to calculate their combined effect on final abundances. After a presentation of the method, results from application to three different nucleosynthesis processes are shown: the $\gamma$-process and the $s$-process in massive stars, and the main $s$-process in AGB stars (preliminary results). Thermal excitation of nuclei in the stellar plasma and the combined action of several reactions increase the final uncertainties above the level of the experimental errors. The total uncertainty, on the other hand, remains within a factor of two even in processes involving a large number of unmeasured rates, with some notable exceptions for nuclides whose production is spread over several stellar layers and for $s$-process branchings.
\end{abstract}

\section{INTRODUCTION}
Nucleosynthesis studies involve following the temporal abundance changes in an astrophysical site in nuclear reaction networks \cite{rauscher11,rauscher14}. The hydrodynamical evolution in the astrophysical site provides the history of changes in matter density and plasma temperature affecting the nuclear reactions proceeding within the stellar plasma. Energy generation by these nuclear reactions, also calculated by the reaction networks, has to be fed back into the hydrodynamic evolution when considering the most general situation. Since the energy generation is governed by comparatively few reactions on lighter target nuclei, many investigations use a simplified approach with two reaction networks to keep computational cost low. A smaller network accounting for the energy generation is coupled directly to the hydrodynamics, whereas a larger reaction network is run in parallel to account for abundance changes of other interesting nuclides. Instead of running in parallel, the larger network can also be run in a post-processing approach, applying the previously calculated density and temperature as functions of time (in the following called trajectory).

The final abundances obtained in such computations carry a combined uncertainty stemming from the uncertainty in the hydrodynamic evolution of a chosen astrophysical site (and in the choice of site) and the uncertainties in the nuclear reaction rates employed in the reaction networks. The latter uncertainties again are either experimental or theoretical uncertainties or a combination. Ideally, for a full error analysis it would be necessary to propagate both uncertainties, astrophysical and nuclear, through the full calculations. Currently, this is not feasible computationally and is difficult also because it is impossible to quantify an error in a choice of model \cite{rauscher14,rauscher16}. Therefore one resorts to studying uncertainties coming from reaction rates separately from astrophysical uncertainties. Even in this case, simplifications are made. While it would be desireable to start from quantified uncertainties in the description of basic nuclear properties and propagate them into the astrophysical reaction rates, this so far has only been attempted for certain reactions on lighter targets with experimentally determined properties \cite{starlib}. Mostly, the dependence of final abundances to single reactions and their uncertainties are investigated by manually varying these single reactions separately in post-processing, assuming more or less arbitrary variation factors.

\section{\textsc{PizBuin} MONTE CARLO CODE}

\begin{figure}[t]
  \centerline{\includegraphics[width=\columnwidth, viewport= 80 30 780 500]{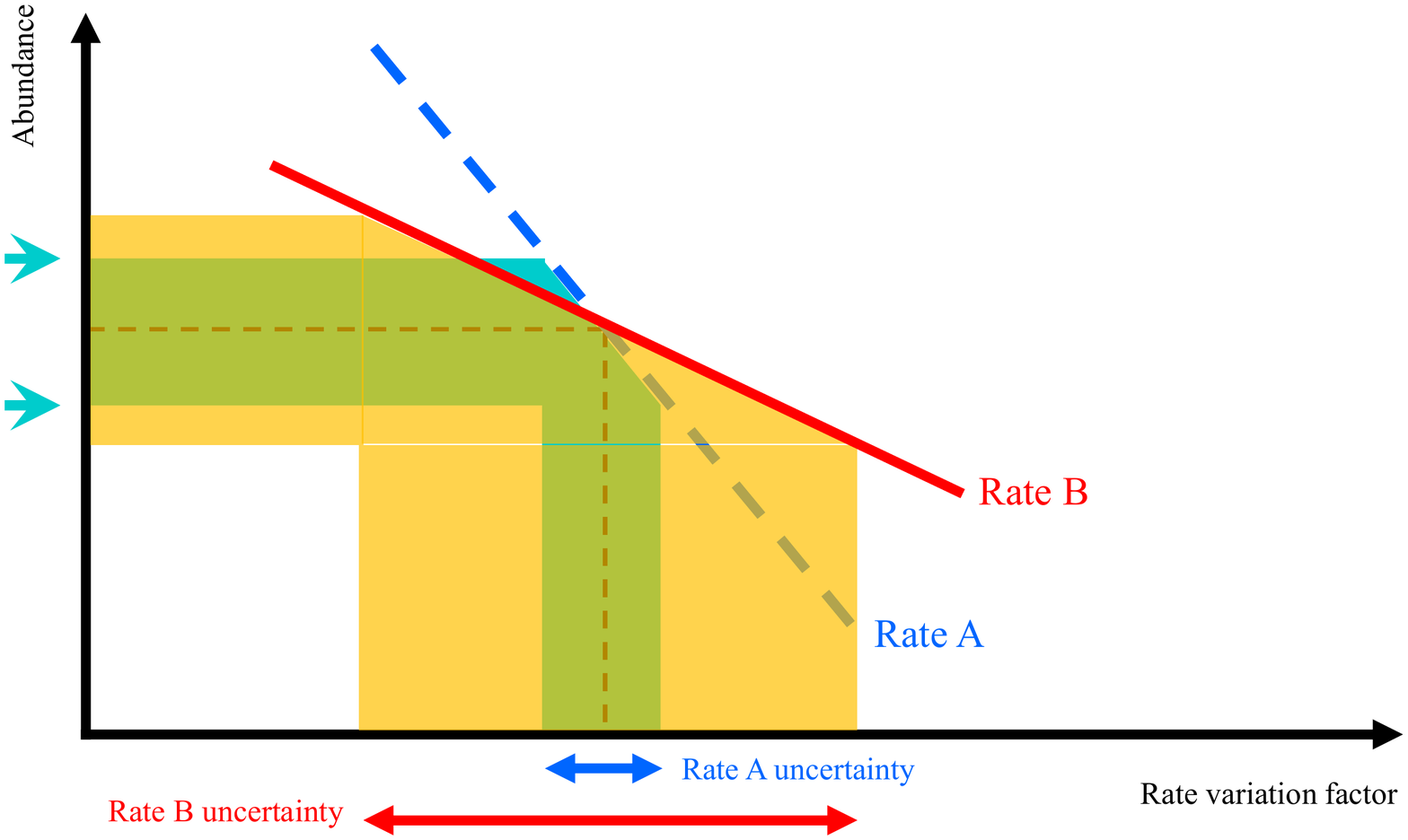}}
  \caption{\label{fig:linedrawing}A simplified sketch of the impact of reaction rate variations on the variation of a final abundance. Two reactions are assumed to contribute and the sensitivity of the final abundance to a change in these rates is given by the slope of the two lines. (Positive slopes are for production reactions, negative ones for reactions reducing the abundance of a nuclide.) Although the abundance is more sensitive to rate A (dashed blue line) with its steeper slope, the abundance uncertainty is dominated by rate B (full red line) because its rate uncertainty is larger than the one of rate A.}
\end{figure}

We improve on the above mentioned, separate variations of individual rates in several ways:
\begin{enumerate}
\item Instead of individual variation of a few selected rates \textit{all} rates in the network (or in a region of the network) are comprehensively varied simultaneously in a Monte Carlo (MC) procedure.
\item Key reactions are identified by inspection of \textit{correlations in the simultaneous variation of all rates} instead of relying on the sensitivity of an abundance to the individual variation of a single rate.
\item Each rate is assigned an individual uncertainty which is \textit{temperature dependent} and which is sampled by a different MC variation factor for each rate. Uncertainties do not have to be symmetric.
\item The bespoke rate uncertainties are derived from a \textit{combination of experimental and theoretical uncertainties} where a measurement of a reaction cross section is available.
\end{enumerate}

Items 1 and 2 in the above list are an important improvement over previous approaches using a decoupled variation of individual rates. Such individual rate variations may be misleading and may result in an incorrect assessment of total uncertainties as well as the importance of the selected rates. This is due to the fact that the combined action of several reactions can cover or enhance uncertainties in each single rate. A simplified, schematic illustration of this is shown in Fig.\ \ref{fig:linedrawing}. (In actual calculations, the dependence of an abundance on a rate will not be linear and more than two reactions may contribute to production and destruction of a nuclide, leading to a more complicated dependence of the final abundance.) This also implies that the importance of a reaction cannot be derived simply from the abundance change $dY$ when varying its reaction rate $\mathcal{R}$, i.e., from the slope $\left|dY/d\mathcal{R}\right|$, as it is often done in nucleosynthesis studies. We rather define a \textit{key reaction} as a reaction dominating the uncertainty of the final abundance. This means that this abundance uncertainty will be considerably reduced when better constraining the corresponding key reaction by experiment or/and theory. The identification of key reactions is performed automatically by calculating correlation factors between variation factors of all reactions and corresponding variations in the final abundances. Thus, key reactions are specific to a nuclide and it is possible that no key reaction can be found for a given nuclide when many reactions are contributing to its abundance. For further details, see \cite{rauscher16}.

The MC procedure was implemented in the \textsc{PizBuin} code suite which includes a parallelized MC driver supplying MC variation factors to a fast reaction network. Tools to analyze the MC output complete the code suite. Typical network sizes included several thousand target nuclei at and around the valley of stability, depending on the investigated nucleosynthesis process. Choosing a sufficiently large number of MC iterations (i.e., rate variations) ensures that the space of possible reaction rate combinations is well sampled. Using test calculations it was confirmed that 10000 MC iterations are sufficient, i.e., each rate is varied 10000 times per trajectory. The required computational time is independent of how many rates are varied but rather depends on the time to complete one network run, the number of MC iterations, and the number of trajectories to be processed. Most processes require more than one trajectory (and up to several thousand) to be considered, one for each zone or tracer. For a complete result, abundance changes in all trajectories have to be included and analyzed together. This necessitates the use of parallel processing to speed up the computations.

\section{REACTION RATE UNCERTAINTIES AND MC VARIATION FACTORS}

An essential ingredient in the MC procedure are the assumed uncertainties for the individual reaction rates. It is important to remember that \textit{stellar} reaction rates not only include reactions proceeding on the ground state (g.s.) or long-lived isomeric state of a target nucleus but also on a distribution of its excited states. For the overwhelming majority of experimentally determined rates (and essentially for all rates beyond the Fe-group) only the g.s.\ contributions to the stellar rate have been determined, if there was a measurement at all at astrophysically relevant energies. Although the number of target nuclei in excited states in the plasma obeys a Boltzmann distribution exponentially declining with excitation energy, a similar Boltzmann factor in the rate definition almost offsets the exponential and turns it into a linear dependence of the excited state contributions \cite{rauscher11}. This yields sizeable excited state contributions even at comparatively low plasma temperatures. Since the cross sections of excited states are mostly unmeasured, a combination of experiment and theory uncertainties is called for. The relative contribution of the g.s.\ to the stellar rate $X_0(T)$ depends on the plasma temperature $T$ \cite{rauscher14,sprocuncert},
\begin{equation}
\label{eq:xfactor}
X_0(T)= \frac{2J_0+1}{G(T)} \frac{\mathcal{R}^\mathrm{g.s.}(T)}{\mathcal{R}^*(T)}\quad ,
\end{equation}
where $\mathcal{R}^*$ is the stellar rate and $\mathcal{R}^\mathrm{g.s.}$ is the rate for the reaction on the g.s.\ of the target nucleus (which usually is the experimental rate). The nuclear partition function is denoted by $G(T)$ and the g.s.\ spin of the target nucleus is $J_0$.
It is very important to note that this definition of the g.s.\ contribution is different from the simple ratio of stellar and g.s.\ rate, as previously erroneously used when a ``stellar enhancement factor'' was given. Exhaustive tables of true g.s.\ contributions as defined in Eq.\ (\ref{eq:xfactor}) are found in \cite{sprocuncert,apjsuppl}.

Obviously, also the resulting combined uncertainty $u^*$ of the stellar rate -- constructed from an experimental uncertainty factor $U_{\mathrm{exp}}$ for the reaction cross section of a target nucleus in the g.s.\ and a theoretical one $U_\mathrm{th}$ for the prediction of reactions with the target nucleus being excited -- is temperature dependent according to the varying contributions of g.s.\ and excited states \cite{rauscher14},
\begin{equation}
u^*(T)=U_\mathrm{exp} +\left( U_\mathrm{th}-U_{\mathrm{exp}}\right) \left(1-X_0 \left(T\right) \right) \label{eq:uncertainty} \quad,
\end{equation}
with $U_\mathrm{exp}<U_\mathrm{th}$. In the absence of any measurement the above equation trivially reduces to just the theory uncertainty. Note that $U_{\mathrm{exp}}$, $U_\mathrm{th}$, and $u^*$ are uncertainty \textit{factors} by which a rate is multiplied or divided to define an uncertainty range. Instead of one factor $u^*$, asymmetric uncertainties can be allowed by using different factors $u^*_\mathrm{lo}(T)$ and $u^*_\mathrm{hi}(T)$ for the lower and upper limit, respectively.
Within an uncertainty range $[r^*_0(T)/u^*_\mathrm{lo}(T),r^*_0(T) u^*_\mathrm{hi}(T)]$ a probability distribution has to be chosen for each reaction to be varied. The \textsc{PizBuin} driver allows to choose among several options, including Gaussian, lognormal, and uniform distribution. For the investigations presented here, we chose to draw values from a uniform distribution within the given limits. For $U_{\mathrm{exp}}$ the experimental $2\sigma$ error of each reaction was taken as upper and lower limit. For the theory uncertainty different, partly asymmetric, uncertainty factors were chosen, depending on the type of reaction (see \cite{rauscher16,nobuya17} for details).

In each MC iteration, the \textsc{PizBuin} driver supplies a vector containing a random value $0<v<1$ for each reaction to be varied. The random value is drawn from the probability distribution chosen for that reaction. In the network calculation it is mapped to an actual variation factor $f(T)$ for each rate according to
\begin{equation}
f(T)=\frac{1}{u^*_\mathrm{lo}} + v \left( u^*_{\mathrm{hi}}-\frac{1}{u^*_{\mathrm{lo}}} \right) \quad . \label{eq:varfact}
\end{equation}
This means that the stellar rate value $r^*_\mathrm{new}(T)$ actually used in the reaction network is given by $r^*_\mathrm{new}(T)=f(T) r^*_0(T)$, where $r^*_0(T)$ is the original, unchanged stellar rate as given in the reaction rate library. Forward and reverse rates are varied by the same factor as they are connected by detailed balance.

The temperature-dependent uncertainty factors $u_\mathrm{weak}^*$ for stellar electron captures, $\beta^+$-, and $\beta^-$-decays were derived slightly differently than described above because the g.s.\ contributions $X_0$ were not known. The theory uncertainty for predicting weak reactions on excited states was set to $U_\mathrm{th}^\mathrm{weak}=10$ and the uncertainty was computed as
\begin{equation}
u_\mathrm{weak} ^* = u_\mathrm{weak,lo} ^* = u_\mathrm{weak,hi} ^* = \frac{\left( 2J_0+1\right) {U_\mathrm{g.s.}^\mathrm{weak}}}{G(T)}+ {U_\mathrm{th}^\mathrm{weak}}\left( 1 - \frac{2J_0+1}{G(T)}  \right) \quad, \label{eq:betauncert}
\end{equation}
where $U_\mathrm{g.s.}^\mathrm{weak}$ is either the uncertainty of a measured half-life or a factor 1.3. Thus, the uncertainty increases with increasing temperature also for these rates.

\section{APPLICATION TO NUCLEOSYNTHESIS PROCESSES}

The MC procedure described in the preceding sections was already applied to a number of nucleosynthesis sites and processes, using the \textsc{PizBuin} code suite. In the following, a brief overview of the results is given. For further details, please refer to the original publications.

The focus of the work presented here is on studying abundance uncertainties from Fe isotopes up to heavier nuclei. Determining abundance uncertainties in this region is interesting because the contributing nucleosynthesis processes involve a large number of unmeasured, predicted rates with possibly large rate uncertainties. Moreover, as shown previously \cite{rauscher14,sprocuncert,apjsuppl}, in this region even measured rates can exhibit an uncertainty which is larger than the experimental one, due to the contributions of reactions on thermally excited states. This is most pronounced at elevated temperature but can already occur at $s$-process temperatures for deformed nuclei \cite{sprocuncert,apjsuppl}. On the other hand, studying this region is simplified by the facts that most reactions (except close to the driplines and some magic nuclei) can be described in the Hauser-Feshbach model \cite{rtk,rath}, simplifying the quantification of uncertainties, and that these reactions' energy production can be neglected, justifying the post-processing approach.

\subsection{$\gamma$-Process in Massive Stars}

\begin{figure}[t]
  \centerline{\includegraphics[width=0.5\columnwidth]{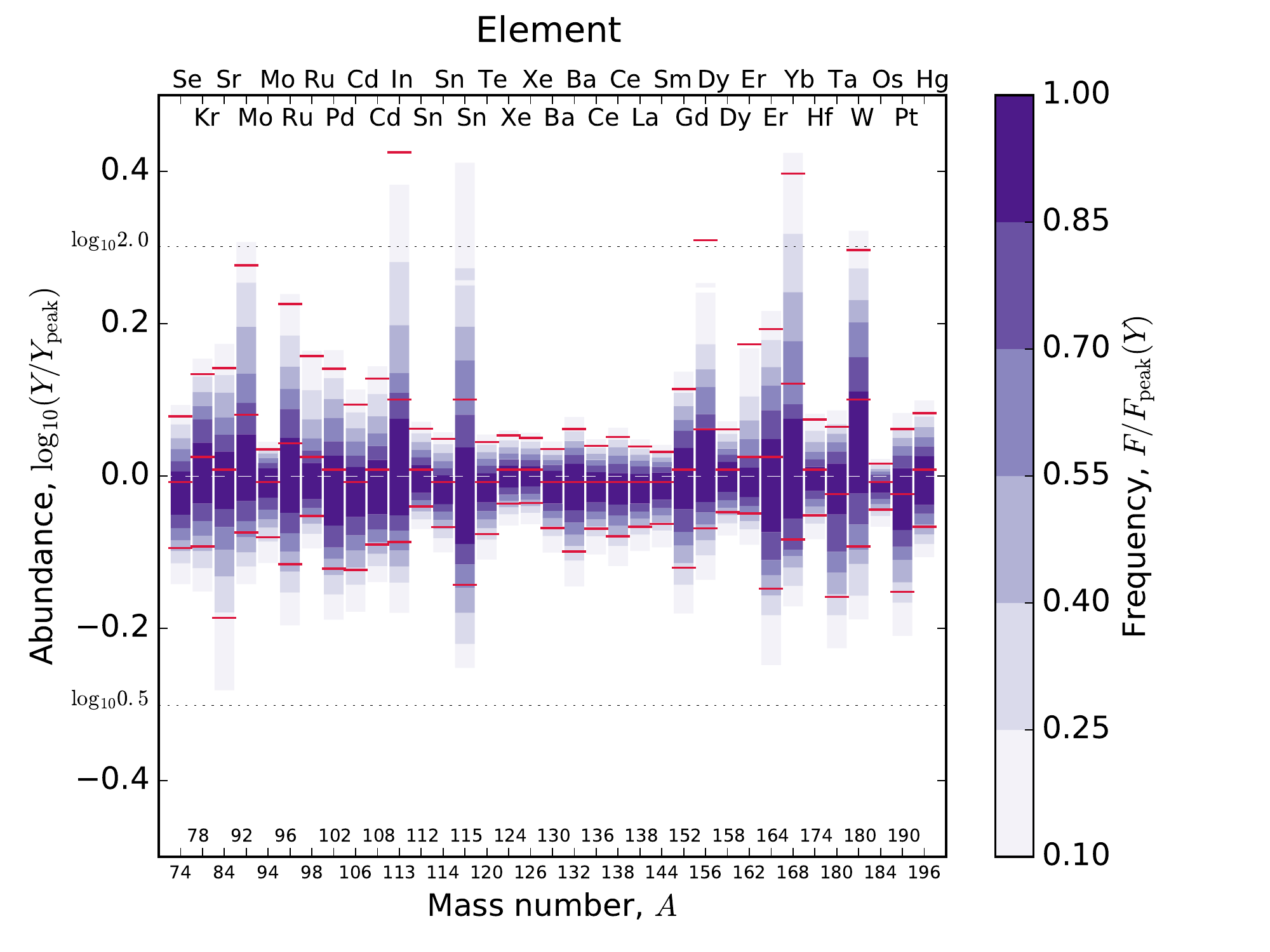}
  \includegraphics[width=0.5\columnwidth]{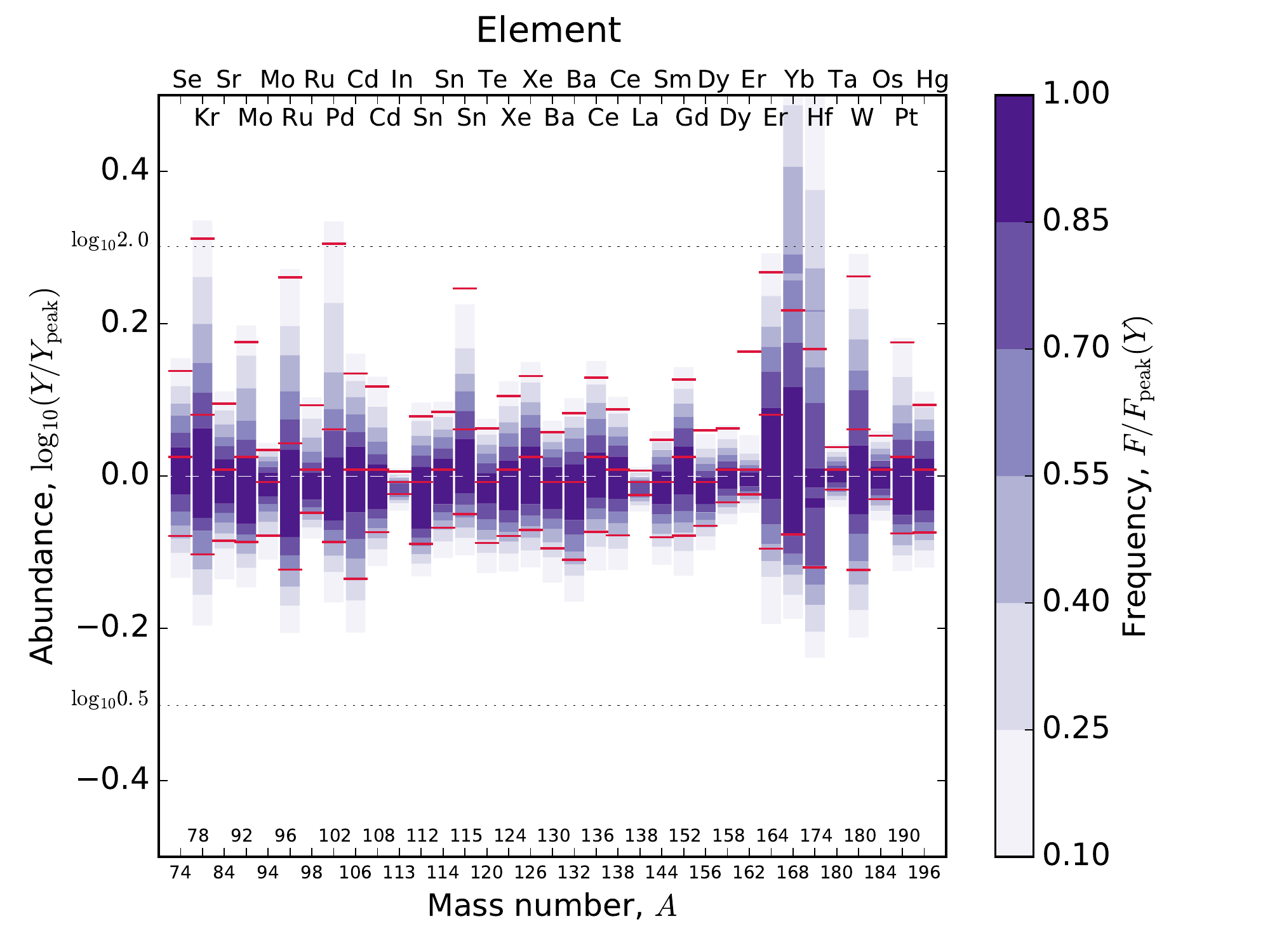}}
  \caption{\label{fig:gammaproc}Final production uncertainties of $p$-nuclei due to rate uncertainties in the $\gamma$-processes of exploding 15 $M_\odot$ (left) and 25 $M_\odot$ (right) stars. The color shade gives the relative probabilistic frequency and the red lines enclose a 90\% probability interval for each nuclide. Note that the uncertainties are asymmetric. Figures are taken from \cite{rauscher16}.}
\end{figure}

The first application of \textsc{PizBuin} was to study the uncertainties in abundances obtained by the $\gamma$-process in massive stars \cite{rauscher16}. Photodisintegration reactions in the O/Ne-shell of a massive star during its supernova explosion produce neutron-deficient isotopes and this mechanism was suggested as a possible source of the so-called $p$ nuclei. These are proton-rich isotopes which can be produced neither in the $s$- nor the $r$-process \cite{arngor,p-review}.
Since the $p$-nucleus production depends on the mass of the supernova progenitor, trajectories and initial abundances obtained in a 15 $M_\odot$ and a 25 $M_\odot$ model evolved with the KEPLER stellar evolution code \cite{rhhw02,hegwoo07} were used. Earlier post-processing studies (e.g., \cite{rapp,reifarth,marco}) with individual variation of reaction rates for the $\gamma$ process used trajectories from the O/Ne shell of another 25 $M_\odot$ model \cite{hashi} and final abundance uncertainties were also obtained with those trajectories for comparison. This allowed not only to disentangle nuclear and astrophysical uncertainties but also to draw some conclusions on differences between astrophysical models.

Figure \ref{fig:gammaproc} shows final abundance uncertainties for the 15 $M_\odot$ and 25 $M_\odot$ KEPLER models. The differences between the two models seen in the figure are due to the different combinations of peak temperature and density attained when the supernova shock is passing through the zones in these models. The overall uncertainties are much smaller than a factor of two despite the fact that most contributing reactions are unmeasured. Larger uncertainties appear in the region from Er to W in both models and for $^{113}$In and $^{115}$Sn in the 15 $M_\odot$ model. For these cases, the production is spread out over many different zones (trajectories), contrary to the other nuclides which are produced in one or only few zones. This implies that they are produced across a larger range of temperatures and densities and that therefore many different reactions contribute. Incidentally, the region around mass number $A\approx 165$ was also identified as a problematic region in \cite{rhhw02}, where the solar system $p$-abundance pattern could not be reproduced. The current findings indicate that a nuclear physics solution to this problem, i.e., through improved reaction rates, may still not be ruled out. The nuclides $^{113}$In and $^{115}$Sn may receive contributions from the $s$- and the $r$-process \cite{nemeth} and if this is the case their larger uncertainty here is inconsequential.

Key reactions were identified for all nuclides except for those produced across many zones. Complete lists are given in \cite{rauscher16}.

A different uncertainty pattern was found using the alternative trajectories from the calculation by \cite{hashi}. The differences were traced back to the different zoning used in that model. The grid of mass zones is much cruder than in the KEPLER models and thus the resolution of peak temperatures is much lower. This does not allow to follow the peak temperatures sufficiently well but rather lumps several peak temperatures into a zone with a single peak temperature and therefore overemphasizes the importance of certain reactions. Furthermore, the available set of trajectories is too restricted and does not include inner zones still contributing to the abundances of light $p$-nuclei.

\subsection{(Weak) $s$-Process in Massive Stars}

\begin{figure}[t]
  \centerline{
  \begin{tabular}{c}
  \includegraphics[width=0.4\columnwidth]{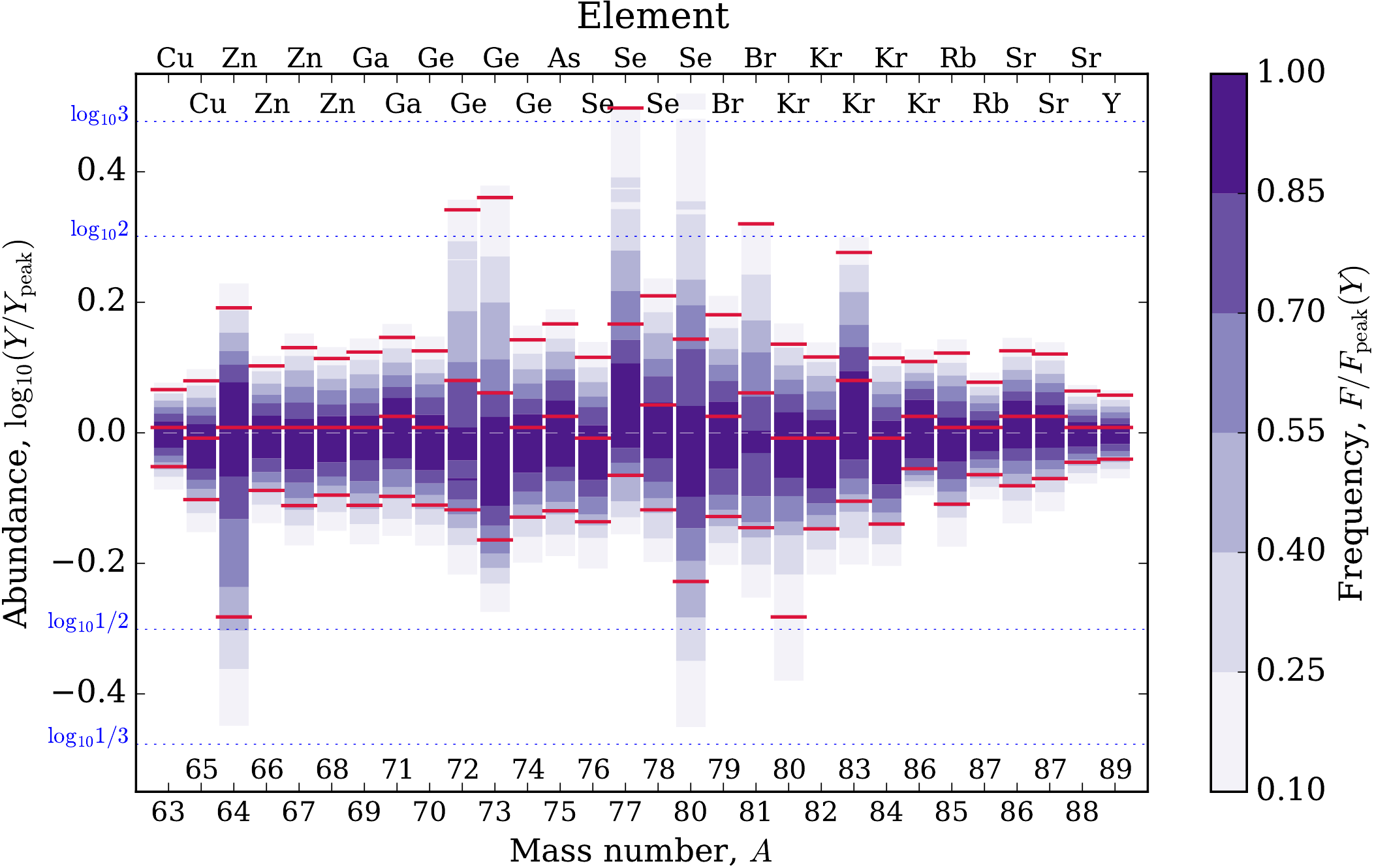}\\
  \includegraphics[width=\columnwidth]{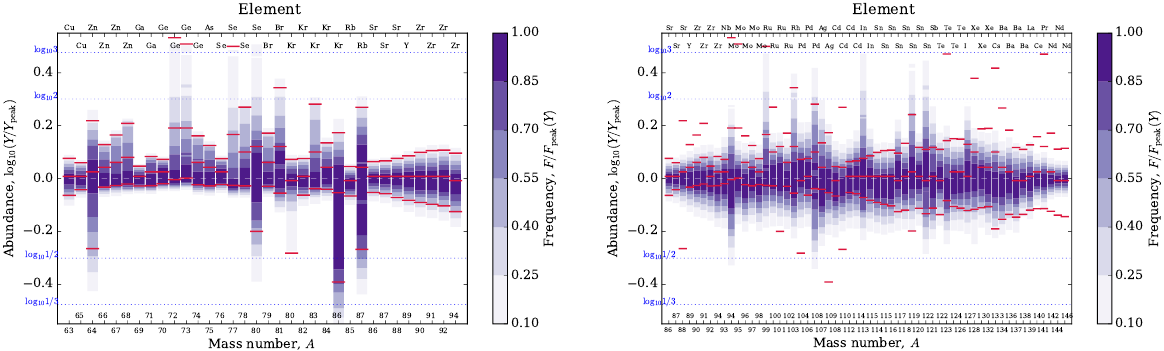}
  \end{tabular}
  }
  \caption{\label{fig:sproc}Same as Fig.\ \ref{fig:gammaproc} but for the $s$-process in a 25 $M_\odot$ star. Uncertainties for a non-rotating, solar metallicity model and a rotating, metal-poor star are shown in the upper and lower panels, respectively. Figures are taken from \cite{nobuya17}.}
\end{figure}

During core He-burning and C-shell burning of massive stars, the reaction $^{22}$Ne($\alpha$,n)$^{25}$Mg acts as a neutron source allowing the production of nuclides up to mass number $A\approx 90$ beyond the Fe-group. This production can be identified as the weak $s$-process component in the classical picture of the $s$-process. The nucleosynthesis path is defined by a sequence of neutron captures and $\beta^-$-decays following the valley of stability in the nuclear chart. Rotation does not affect the path in models with solar metallicity. At lower metallicity, however, rotation-induced mixing of parts of the He-burning core and the H-burning shell leads to a considerable enhancement in the neutron production. When $^{12}$C and $^{16}$O are transferred from the He-core to the H-shell, $^{14}$N is produced via CNO cycles. This additional $^{14}$N can be mixed back into the core where it is converted to $^{22}$Ne which then acts as an additional neutron source in the classical sense. The available larger neutron density at the end of core He-burning together with the lower seed abundance in a metal-poor star allows a considerable enhancement in the production of nuclei with mass numbers $A>100$, even up to Ba \cite{nobuya17,frischknecht12,frischknecht16}. This is a possible source of heavy elements in the early Galaxy and can have an important impact on chemical evolution.

Production uncertainties of the regular weak $s$-process in a 25 $M_\odot$ star with solar metallicity were studied along with the $s$-process in a 25 $M_\odot$ rotating, metal-poor star \cite{nobuya17}. Figure \ref{fig:sproc} provides an overview of the results for both models. As expected, the non-rotating model at solar metallicity only exhibited a weak $s$-process, whereas the rotating, low-metallicity model produced significant amounts of nuclides up to Ba. Due to the larger neutron density in the rotating model, the nucleosynthesis path is slightly shifted to the neutron-rich side compared to the non-rotating model and therefore shows a different uncertainty pattern also in the range of lighter nuclides (bottom left panel of Fig.\ \ref{fig:sproc}). In the solar metallicity model, uncertainties are symmetric with few exceptions, also showing larger uncertainties. Apart from these exceptions, uncertainties increase with increasing mass number, reflecting the propagation of uncertainties with the nucleosynthesis flow. Above mass number $A=80$ the production falls off sharply and since only experimentally well-constrained rates are encountered the final uncertainties are reduced again. In the rotating, low-metallicity model, on the other hand, uncertainties are strongly asymmetric and the propagation of increasing uncertainties is continued also beyond Sr. Only after Ba, production uncertainties are suppressed again as the nuclear flow is fading.

Neutron capture uncertainties dominate the final abundance uncertainties in both models, except for the branchings determining the abundances of $^{64}$Zn, $^{80}$Se (both models) and $^{94}$Nb, $^{108}$Pd, $^{122}$Sn (rotating model only) where the $\beta^-$-decay uncertainty becomes important. The final uncertainties generally are larger than the experimental errors on the individual neutron capture rates. This underlines the importance of the contribution of thermally excited states and also the increase in uncertainties through the combined action of several reactions and its propagation.

Key reactions were identified in the non-rotating, solar metallicity model and the rotating, low-metallicity model. Complete lists are given in \cite{nobuya17}.

\newpage

\subsection{Main $s$-Process in Thermonuclear Pulses of AGB Stars}

\begin{figure}[t]
  \centerline{\includegraphics[width=\columnwidth,viewport=150 130 1560 830]{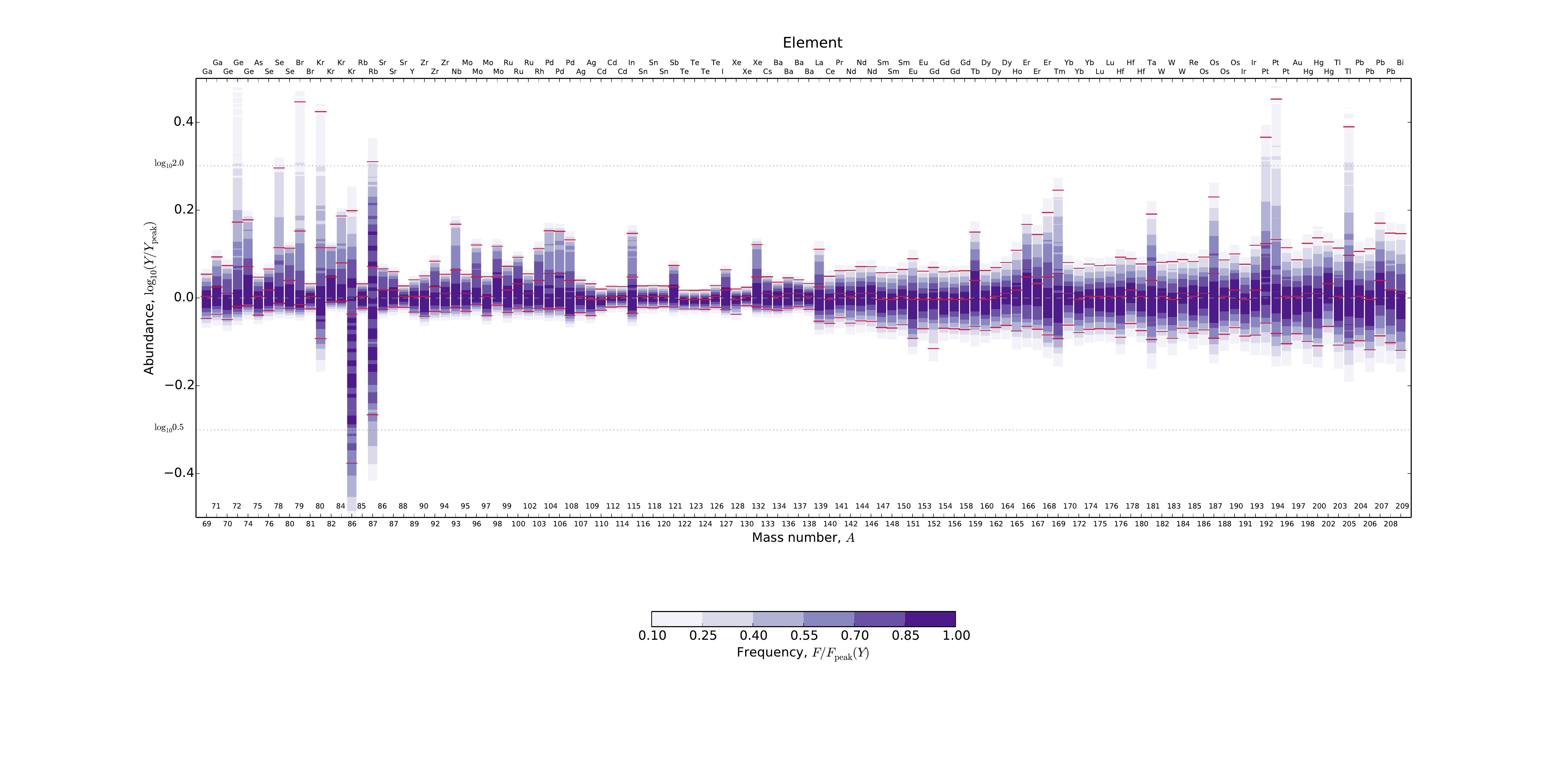}}
  \caption{\label{fig:mainsproc}Same as Fig.\ \ref{fig:gammaproc} but for the main $s$-process component from an AGB star.}
\end{figure}

The main component of the $s$-process, accounting for approximately half of the stable nuclei above mass number $A \approx 90$, was found to be produced in thermal pulses of stars with masses below 8 $M_\odot$, the so-called AGB stars. The thermal pulses mix protons from the outer H-layer into the He-layer below. This causes the production of $^{13}$C, enabling the very efficient neutron source  $^{13}$C($\alpha$,n)$^{16}$O which, in turn, ensures neutron irradiation of material in the He-shell over the long time between and during He-shell burning \cite{arl}.

Here, preliminary results on abundance uncertainties caused by nuclear reactions in this process are presented.
Although a detailed treatment of the mixing mechanism would require the most advanced 3D models, for our purpose we preferred to use a simple
trajectory describing a single thermal pulse in the He-shell, assuming the mixing to have taken place already before. Although abundances are increased with time, the nucleosynthesis path is not changing with the number of pulses and similar uncertainties are expected in each pulse. For this study, a 1D trajectory of a solar metallicity model calculated by the MESA code \cite{MESA} was used. The chosen 3 $M_\odot$ model is considered to dominantly contribute to chemical evolution and it was checked that the trajectory reproduces the typical abundance pattern of the main $s$-process component.

A preliminary view of the resulting production uncertainties is shown in Fig.\ \ref{fig:mainsproc}. A detailed analysis and discussion will be provided in a forthcoming paper \cite{cescutti}. Similar arguments as given above for the $s$-process in massive stars apply also here, concerning the size of the uncertainties and the propagation effect which can be seen especially well above Ba. The impact of $\beta^-$-decay uncertainties at branchings is expected to be stronger than in massive stars because the achieved neutron densities are lower, slowing down neutron captures with respect to $\beta^-$-decays.

\section{SUMMARY AND OUTLOOK}

A Monte Carlo procedure to systematically derive final abundance uncertainties stemming from uncertainties in astrophysical reaction rates was presented. This procedure is implemented in the \textsc{PizBuin} code suite. Since it is based on a simultaneous, systematic variation of a large number of rates within individual rate uncertainties, it captures the effect of a concerted action of several rates which cannot be treated by an individual variation of isolated rates. For the first time, temperature-dependent uncertainties were introduced. Key reactions are determined from correlations obtained from combined MC data of runs with all available trajectories instead of simply using the sensitivity of an abundance to a rate. This method is considered superior to individual rate variations and inspection of flow plots, and supersedes previous approaches. Thus, the obtained results provide a sound basis for further investigations in Galactic Chemical Evolution and also -- after careful consultation of g.s.\ contributions in key rates and absolute production amounts of nuclides -- for planning experimental efforts to better constrain key reactions.

Applications of the MC procedure to the $\gamma$- and $s$-process in massive stars, and the main $s$-process in an AGB model were presented. Currently, investigations of further nucleosynthesis processes are in progress, such as the $\gamma$-process in SN Ia, the $\nu p$-, and the $r$-process.

\section{ACKNOWLEDGMENTS}
This project has been financially supported by the ERC
(EU-FP7-ERC-2012-St Grant 306901, EU-FP7 Adv Grant GA321263-FISH), the EU COST Action CA16117 (ChETEC), and
the UK STFC (ST/M000958/1). Parts of computations were carried out by
COSMOS (STFC DiRAC Facility) at DAMTP in University of Cambridge.
G.C.\ acknowledges financial support
from the EU Horizon2020 programme under the Marie Sk\l odowska-Curie
grant 664931.
%

\end{document}